\documentclass[pre,twocolumn,floats]{revtex4}

\usepackage{amsmath}
\usepackage[dvips]{graphicx}

\begin{document}

\title{Sharp interface limit of a phase-field model of crystal grains}

\author{Alexander E.~Lobkovsky} \author{James A.~Warren}
\affiliation{National Institute of Standards and Technology, 100
  Bureau Drive, Gaithersburg, MD, 20899}

\begin{abstract}
  We analyze a two-dimensional phase field model designed to describe
  the dynamics of crystalline grains.  The phenomenological free
  energy is a functional of two order parameters. The first one
  reflects the orientational order while the second reflects the
  predominant local orientation of the crystal.  We consider the
  gradient flow of this free energy.  Solutions can be interpreted as
  ensembles of grains (in which the phase of the order parameter is
  approximately constant in space) separated by grain boundaries.  We
  study the dynamics of the boundaries as well as the rotation of the
  grains.  In the limit of the infinitely sharp interface, the normal
  velocity of the boundary is proportional to both its curvature and
  its energy.  We obtain explicit formulas for the interfacial energy
  and mobility and study their behavior in the limit of a small
  misorientation.  We calculate the rate of rotation of a grain in the
  sharp interface limit and find that it depends sensitively on the
  choice of the model.
\end{abstract}
\pacs{81.10.Aj, 81.30.-t, 81.10.Jt, 61.72.Mm}

\maketitle

\section{Introduction}
\label{sec:intro}

The characterization and evolution of microstructure forms a
cornerstone of materials science.  In particular the grain structure
of a polycrystalline material determines many of its properties.
Recent efforts at modeling the evolution of grain boundaries have used
a variety of approaches \cite{Morin}, \cite{Lusk}, \cite{Chen},
\cite{Holm}.  Herein we focus on the recently introduced phase field
model of Kobayashi, Warren and Carter (KWC)\cite{kwc3} which is based
on earlier attempts by the same authors (\cite{kwc} and \cite{wkc}).

The model is motivated by symmetry principles and has a surprisingly
rich set of physical characteristics.  In particular, KWC showed
numerically that solutions to this model can be interpreted as a
collection of grains.  The velocity of the interface between the
grains was found to be approximately proportional to the local
curvature of the interface, but the grains were also able to rotate
towards lower energy misorientations.  In support of the notion of
grain rotation, three independent unpublished molecular dynamics
studies by M.~Upmanyu and D.~Srolovitz, S.~R.~Phillpot and D.~Wolf,
and S.~Srivillaputhur and J.~W.~Cahn, suggest that grain rotation will
occur under certain circumstances.  In addition, there is a
substantial history (and debate) concerning the mechanisms \cite{Li},
\cite{Martin} and observation of grain rotation
\cite{King,chen_balluffi}.

The KWC phase field model is challenging mathematically because of a
singular term in the free energy.  Kobayashi and Giga \cite{kg99}
studied similar singular models and showed that there is a way to
handle the singularity consistently.  In this paper we will apply
their method to the KWC model and show that its solutions can indeed
be interpreted as grains.  Our main goal is to analytically obtain the
properties of a grain boundary as well as the rotation rate of a
grain.  We accomplish this task by considering a distinguished limit
of the model parameters in which the width of the boundary vanishes
while its measurable characteristics remain finite and non-zero.  The
methodology of this this so called sharp interface limit is well
established \cite{Caginalp}, \cite{Mcfadden}, \cite{fife}.

The organization of this paper is as follows.  In Sec.~\ref{sec:model}
we introduce the order parameters, phenomenological free energy, and
the gradient flow equations.  We next perform a formal asymptotic
expansion of the model in Sec.~\ref{sec:formal_asymptotics}.  The
zeroth order probelm of this expansion is discussed in
Sec.~\ref{sec:zeroth}, where we obtain the profile and the energy of
the static flat boundary.  In the following section
(Sec.~\ref{sec:first}) the first order asymptotics are examined.  We
are able to determine the velocity of a curved grain boundary.  We
find that this velocity is proportional to both the curvature and the
interfacial energy and obtain an expression for the mobility of the
interface.  As noted above, grains can rotate in this model.  We
calculate the rate of rotation of a grain in Sec.~\ref{sec:rotation}.
Finally, in Sec.~\ref{sec:toy} we apply all of the results described
herein to the simple case of a circular grain embedded in a matrix.
In our conclusion we discuss further ideas and the implications of
this phase field model for the problem of coarsening in polycrystals
in Sec.~\ref{sec:discussion}.

\section{Model}
\label{sec:model}

We model the evolution of a collection of nearly perfect crystalline
grains in two dimensionsvia a phase field model.  First, we discuss
order parameters which capture the microscopic physics of grain
boundaries.  It is then possible to construct a phenomenological free
energy which favors a perfect uniform crystal and supports stable
grain boundaries.  Evolution of an ensemble of grains is then modeled
via gradient flow of this free energy.

Following \cite{kwc3} we develop two order parameters which capture
the physics of grain boundaries.  To distinguish grains of different
orientations we introduce a continuously varying local orientation
$\theta$.  Since the energy of crystal does not depend on $\theta$
itself, the phenomenological free energy will be a functional of the
gradients of $\theta$ only.  The second order parameter $\eta$ is used
to differentiate the nearly perfect crystal in the interior of the
grains from the disordered material in the grain boundary.  It varies
from perfect order $\eta = 1$ to complete disorder $\eta = 0$.  Both
order parameters are not conserved.

We analyze the free energy (based on KWC)
\begin{multline}
  \label{eq:free_energy}
  {\mathcal{F}}[\eta,\theta] = \\ \frac{1}{\epsilon}\int_\Omega dA
  \left[\frac{\alpha^2}{2} |\nabla \eta|^2 + f(\eta) + g(\eta) \,
    s|\nabla \theta| +  h(\eta) \, \frac{\epsilon^2}{2} |\nabla
    \theta|^2\right],
\end{multline}
where $\alpha$, $\epsilon$ and $s$ are positive model parameters.
Some readers may find the overall prefactor of $1/\epsilon$
suggestive.  Subsequently we will examine the limit $\epsilon
\rightarrow 0$.  This prefactor ensures that the surface energy of a
grain boundary tends to a non-zero constant.  

Term by term, the above free energy deserves some discussion, although
the reader interested in the full motivation of this model is referred
to \cite{kwc3}.  The first term describes the penalty for gradients in
the order parameter (grain boundaries cost energy).  The free energy
density $f(\eta)$ is chosen to be a single well with the minimum at
$\eta = 1$ and $f(1) = 0$ reflecting the fact that disordered material
has higher free energy.  The third and fourth terms are an expansion
in $|\nabla\theta|$, where the couplings $g(\eta)$ and $h(\eta)$ must
be positive definite.  KWC argued that the expansion in
$|\nabla\theta|$ must begin at first order to assure the existence of
stable grain boundaries.  This term yields a singularity in the
dynamic equations.

Indeed, Ref.~\cite{kwc3} omitted the $|\nabla \theta|^2$ term in their
analytical investigation of a stationary flat interface.  This term
was added in \cite{kwc}, for practical reasons, in order to solve the
model numerically.  We shall see that this term makes no qualitative
difference in the properties of static grain boundaries.  However, as
shown in Sec.~\ref{sec:first_small_h}, the mobility of a grain
boundary vanishes as $h^{1/2}$ when in the $h \rightarrow 0$ limit.
Thus, the extra term introduced in KWC seems to be {\em required} for
grain boundaries to move.

Assuming relaxational dynamics for a non-conserved set of order
parameters, we find the gradient flow equations:
\begin{subequations}
  \label{eq:model}
  \begin{eqnarray}
    \label{eq:eta_eq}
    Q(\eta, \nabla \theta)\,\tau_\eta \, \frac{\partial \eta}
    {\partial t} &=& -\epsilon \frac{\delta {\mathcal{F}}}{\delta 
      \eta} \\ \nonumber &=& \alpha^2 \nabla^2 \eta - f_\eta - g_\eta
    \, s |\nabla\theta| - h_\eta \, \frac{\epsilon^2}{2}
    |\nabla\theta|^2, \\ \label{eq:theta_eq}
    P(\eta, \nabla\theta) \, \tau_\theta \, \eta^2 \, \frac{\partial
      \theta}{\partial t} &=& -\epsilon \frac{\delta {\mathcal{F}}}{
      \delta\theta} \\ \nonumber &=& \nabla \cdot
    \left[
      h \, \epsilon^2 \nabla \theta + g \, s\,\frac{\nabla\theta}
      {|\nabla\theta|} 
    \right],
  \end{eqnarray}
\end{subequations}
where we used the subscript $\eta$ to denote differentiation.  The
mobility functions $P$ and $Q$ must be positive definite, continuous
at $\nabla\theta=0$, but are otherwise unrestricted.  The system
(\ref{eq:model}) must be supplemented by initial and boundary
conditions and a rule that specifies the handling of the
indeterminancy (and sigular divergence) of the term
$\nabla\theta/|\nabla\theta|$.  This particular type of the
singularity is generally handled in a theory of extended gradients.
Ref.~\cite{kg99} proves that there is indeed a unique way to prescribe
the value of the right hand side of Eq.~(\ref{eq:theta_eq}) when
$\nabla\theta = 0$.  While we do not wish to attempt to explain all of
the details of Ref.~\cite{kg99}, it is useful to summarize the
ultimate conclusion of the mathematical analysis.  Thus, let us define
a collection $\{G_i\}$ of distinct connected regions where $\nabla
\theta = 0$.  We shall refer to $G_i$ as the interior of grain $i$.
The essence of the method of dealing with the singularity is that the
orientation $\theta$ must remain uniform in space in each $G_i$.
Therefore, the right hand side of Eq.~(\ref{eq:theta_eq}) is chosen in
to be uniform in $G_i$.  This condition, along with the requirement
that $\eta$, $\theta$ and $\nabla \theta$ be continuous at the
boundaries of $G_i$, uniquely determines the rotation rate of each
grain (i.e.\ $\partial \theta/\partial t$ in each $G_i$) and the
motion of its boundaries.
\begin{figure}[htbp]
  \begin{center}
    \includegraphics[width=2.5in]{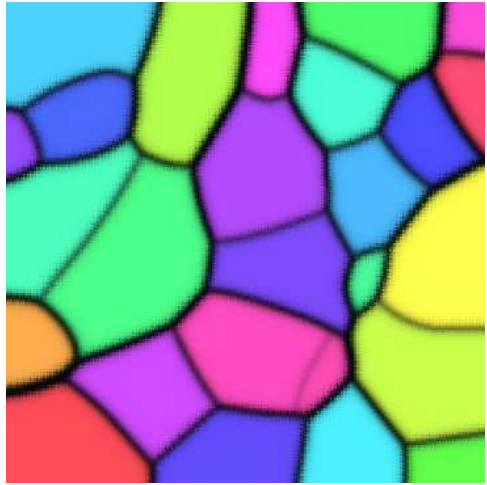}
    \vspace{0.1in}
    \caption{Two-dimensional simulation of Eqs.~(\ref{eq:model}) with
      simple choices for the couplings.  The orientation $\theta$ is
      constant in the shaded grain regions.  The color is used to
      represent the value of $\theta$.  Dark regions correspond to
      small values of $\eta$.}
    \label{fig:grains}
  \end{center}
\end{figure}

Ref.~\cite{2d_sim} verifies numerically that typical solutions of the
gradient flow system (\ref{eq:model}) indeed consist of a collection
of connected regions $G_i$ in which $\nabla \theta = 0$ and $\eta
\approx 1$ (grain interiors) separated by narrow regions (grain
boundaries) in which the orientation $\theta$ changes smoothly between
the neighboring grains.  Thus, the grains rotate ($\theta$ changes in
time uniformly in each grain), and grain boundaries migrate (see
Fig.~\ref{fig:grains} adopted from Ref.~\cite{2d_sim}).

\section{Formal asymptotics}
\label{sec:formal_asymptotics}

\begin{figure}[htbp]
  \begin{center}
    \includegraphics[width=2.5in]{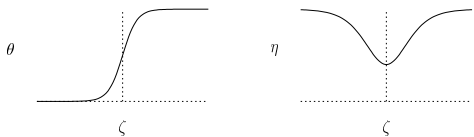}
    \caption{A schematic representation of the solution for a flat
      static interface between two grains.  $\zeta$ is the coordinate
      normal to the interface.}
    \label{fig:1d_bdry}
  \end{center}
\end{figure}

Having elucidated the general properties of the model, we now examine
the sharp-interface limit, in order to extract the physical quantities
from our phase field model.  As discussed earlier, solutions to the
extended gradient flow system (\ref{eq:model}) consist of grains in
which $\nabla\theta = 0$ separated by a network grain boundaries in
which $\nabla\theta \neq 0$.  Although no rigorous proof of the
inevitability of this solution structure exists, there are two
plausibility arguments.  First, once this structure develops, one can
show that it will persist \cite{kg99}.  And second, such a solution
can be explicitly found for a flat stationary interface between two
grains shown schematically in Fig.~\ref{fig:1d_bdry}.  This case
corresponds to the zeroth order of the formal asymptotic expansion as
we show in what follows.

Fig.~\ref{fig:1d_bdry} is highly suggestive.  We have a bicyrstalline
system with a localized region where properties change (i.e.~the
grain boundary).  Let us therefore consider the $\epsilon \rightarrow
0$ limit of the gradient flow system (\ref{eq:model}).  We expect the
width of the grain boundary region, defined by $\nabla \theta \neq 0$,
to shrink to zero.  We employ the standard method of matched
asymptotics \cite{fife} which uses the small ratio of the interface
width to its radius of curvature as an expansion parameter.  In this
limit, all functions and their derivatives vary slowly along the
interface compared to their variation across the interface.  The
analysis can therefore be advanced by defining a coordinate normal to
the interface and scaling it by $\epsilon$.  The order parameters as
functions of this new coordinate are termed the inner solution.  They
are expanded into a power series in $\epsilon$ and the resulting
hierarchy of equations is solved order by order, subject to a matching
condition with the outer solution.  This outer solution, obtained by
setting $\epsilon$ to zero, is valid far away from the interface.

\begin{figure}[htbp]
  \begin{center}
    \includegraphics[width=2.5in]{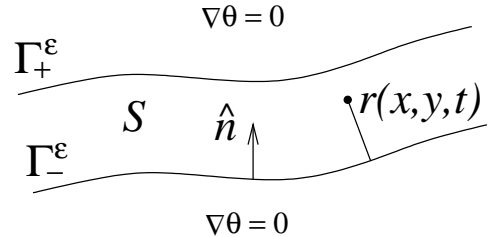}
    \caption{The region $S$ between the curves $\Gamma_-^\epsilon$ and
      $\Gamma_+^\epsilon$ in which $\nabla\theta \neq 0$.}
    \label{fig:bdry}
  \end{center}
\end{figure}

The mathematics of the sharp interface limit of our model is atypical
because the fields $\eta$ and $\theta$ obey two different sets of
equations \cite{kg99}.  Eqs.~(\ref{eq:model}) hold in the grain
boundary region which is defined as a strip $S$ between two smooth
non-intersecting curves $\Gamma^\epsilon_+(t)$ and
$\Gamma^\epsilon_-(t)$ (see Fig.~\ref{fig:bdry}).  Outside this strip
$\nabla \theta = 0$ and
\begin{equation}
  \label{eq:eta_grains}
  Q \,\tau_\eta \, \frac{\partial \eta}{\partial t} = \alpha^2
  \nabla^2 \eta - f_\eta.
\end{equation}

Following \cite{cahn_fife}, let us adopt curvilinear orthogonal
coordinate system $\{r(x, y, t), \, \sigma(x, y, t)\}$.  Let $r(x, y,
t)$ be the distance of the point $(x, y)$ in $S$ from
$\Gamma_-^\epsilon$.  On $\Gamma_-^\epsilon$, coordinate $\sigma$ is
the arc length.  We introduce a scaled variable $\zeta = r/\epsilon$,
and expand $\eta$ and $\theta$ in a formal power series in $\epsilon$
\begin{subequations}
  \label{eq:expansion}
  \begin{eqnarray}
    \label{eq:expansion_eta}
    \eta(\zeta, \sigma, t) &=& \eta_0(\zeta, \sigma, t) + \epsilon \,
    \eta_1(\zeta, \sigma, t) + \ldots, \\ \label{eq:expansion_theta} 
    \theta(\zeta, \sigma, t) &=& \theta_0(\zeta, \sigma, t) + \epsilon
    \, \theta_1(\zeta, \sigma, t) + \ldots. 
  \end{eqnarray}
\end{subequations}
This expansion is valid in $S$ and its immediate neighborhood.  Also
\cite{fife}
\begin{subequations}
  \label{eq:derivatives}
  \begin{eqnarray}
    \nabla &=& \hat n \, \frac{1}{\epsilon} \frac{\partial}{\partial
      \zeta} + \hat t \, \frac{\partial}{\partial \sigma}, \\
    \nabla^2 &=& \frac{1}{\epsilon^2} \frac{\partial^2}{\partial
      \zeta^2} + \frac{\kappa}{\epsilon}
    \frac{\partial}{\partial \zeta} + {\mathcal{O}}(1), \\
    \frac{\partial}{\partial t} &=& -\frac{v}{\epsilon}
    \frac{\partial}{\partial \zeta} + {\mathcal{O}}(1),
  \end{eqnarray}
\end{subequations}
where $\hat n$ is the unit vector normal to $\Gamma^\epsilon_-$
pointing into $S$ ($\zeta$ increases in the direction of $\hat n$) and
$\hat t \perp \hat n$ is the unit vector along the lines of constant
$r$.  The curvature $\kappa = \nabla^2 r$ is positive when $\hat n$
points away from the center of curvature of $\Gamma^\epsilon_-$.
Normal velocity $v = -\partial r/\partial t$ is positive when the
interface moves in the direction of $\hat n$.  This configuration is
shown in Fig.~\ref{fig:bdry}.

In order to proceed we must fix the scaling of $\alpha$, $s$,
$\tau_{\eta}$ and $\tau_\theta$ in the sharp interface limit.  We
select the scaling for which 1) a flat interface does not move, and 2)
all terms in the free energy Eq.~(\ref{eq:free_energy}) scale with the
same power of $\epsilon$.  We also assume that the mobility functions
$P$ and $Q$ are independent of $\epsilon$ in the sharp interface
limit.  These conditions fix
\begin{equation}
  \label{eq:parameter_scaling}
  \alpha = \epsilon \, \tilde \alpha, \quad s = \epsilon \, \tilde s,
  \quad \tau_\eta = \epsilon^2 \tilde \tau_\eta, \quad \tau_\theta =
  \epsilon^2 \tilde \tau_\theta.
\end{equation}

The final ingredient of the formal asymptotic analysis is the matching
of the inner and outer solutions.  The inner solution is found
separately in the grain boundary $S$ ($\nabla\theta \neq 0$) and in
its immediate neighborhood ($\nabla\theta = 0$).  The inner solution
valid in $S$ is denoted by $\eta(\zeta, \sigma, t)$ and $\theta(\zeta,
\sigma, t)$ without superscripts.  The piece of the inner solution
valid in the interior of a grain just outside of $S$ is denoted by
$\eta^{(i)}(\zeta, \sigma, t)$ and $\theta^{(i)}(\zeta, \sigma, t)$.
It is matched with the outer solution $\eta = 1$ and $\theta = const$
for $\zeta \rightarrow \infty$.  We must also match the two pieces of
the inner solution and their derivatives at the boundaries
$\Gamma^\epsilon_\pm$ of the strip $S$ which are at $\zeta =
\zeta_\pm$.

\section{Zeroth order solutions: interface width and energy}
\label{sec:zeroth}

Now, having detailed the formal method of asymptotic expansion, we
proceed to examine the results of this expansion, term by term.  We
begin, naturally enough, with the zeroth order.  The ultimate result
of the zeroth order calculation with be a determination of (i) the
interface width (ii) the surface energy and (iii) the value of the
order parameter $\eta$ in grain boundary.  

Without loss of generality
we can shift the origin of $\zeta$ so that it is in the middle of $S$,
$\zeta_\pm = \pm(\zeta_0 + \epsilon \zeta_1 + \ldots)$.  We first look
at the grain interior region $\zeta > \zeta_+$.  Substituting the
scaling ansatz Eq.~(\ref{eq:parameter_scaling}) and the
$\epsilon$-expansion (\ref{eq:expansion}) into (\ref{eq:model}) and
using (\ref{eq:derivatives}) we obtain
\begin{equation}
  \label{eq:zeroth^m}
  0 = \tilde \alpha^2 (\eta^{(i)}_0)'' - f_\eta(\eta^{(i)}_0), \quad
  (\theta^{(i)}_0)' = 0.
\end{equation}
These solutions must be matched with the outer solution $\eta = 1$ and
$\theta = \theta_+$, as $\zeta \rightarrow +\infty$.  Since the zeroth
order functions $\eta_0$ and $\theta_0$ are independent of $\sigma$,
the coordinate along the interface, and time $t$, they describe a flat
stationary interface.  They should therefore be symmetric with respect
to the center of the boundary $\zeta = 0$.  We can thus set $\theta_+
= \Delta\theta/2$, half the total misorientation (which is in general
a function of time).  Using the fact that $f(1) = 0$ we arrive at
\begin{equation}
  \label{eq:(i)}
  (\eta_0^{(i)})' = \frac{\sqrt{2f(\eta_0^{(i)})}}{\tilde \alpha},
  \quad \theta_0^{(i)} = \frac{\Delta\theta}{2}.
\end{equation}

After similar manipulations we obtain the equations valid in $S$ for
$\zeta \in [-\zeta_+, \zeta_+]$
\begin{subequations}
  \label{eq:zeroth}
  \begin{eqnarray}
    \label{eq:zeroth_eta}
    0 & = & \tilde \alpha^2 \eta''_0 - f_\eta(\eta_0) - \frac{1}{2} 
    h_\eta(\eta_0) (\theta_0')^2 - \tilde s \, g_\eta(\eta_0)
    \theta'_0, \\ \label{eq:zeroth_theta}
    0 & = & [h(\eta_0) \theta'_0 + \tilde s \, g(\eta_0)]',
  \end{eqnarray}
\end{subequations}
Note that we assumed that $\theta' > 0$ in $S$.  This assumption
proves to be unrestrictive since all measurable quantities depend on
the square of this derivative.

To obtain the boundary conditions for $\eta_0$ and $\theta_0$ we
employ two ideas.  First, due to the aforementioned symmetry, $\eta_0$
is even and $\theta_0$ is odd in $\zeta$.  Second, the requirement that
$\eta$, $\theta$ and their derivatives be continuous at $\zeta =
\zeta_+$ can be shown to lead to the continuity of all terms in the
$\epsilon$-expansion (\ref{eq:expansion}) at $\zeta = \zeta_0$.  We
thus obtain
\begin{gather}
  \label{eq:zeroth_eta_bc}
  \eta_0'(0) = 0, \quad \eta_0'(\zeta_0) =
  \frac{\sqrt{2f(\eta_\mathrm{max}})}{\tilde\alpha}, \\
  \label{eq:zeroth_theta_bc} 
  \theta_0(0) = 0, \quad \theta_0(\zeta_0) = \frac{\Delta\theta}{2},
  \quad \theta_0'(\zeta_0) = 0,
\end{gather}
where $\eta_\mathrm{max} \equiv \eta_0(\zeta_0)$.  Let us also
introduce $\eta_\mathrm{min} \equiv \eta_0(0)$.  This designation
reflects our assumption that $\eta_0' > 0$ in $\zeta \in [0, \zeta_0]$
so that $\eta_\mathrm{min} \le \eta \le \eta_\mathrm{max}$.  We can
prove that, in particular cases, this assumption is indeed justified.

Using the last boundary condition we integrate (\ref{eq:zeroth_theta})
to obtain
\begin{equation}
  \label{eq:theta_0'}
  \theta_0' = \tilde s\,\frac{g(\eta_\mathrm{max}) - g(\eta_0)}
  {h(\eta_0)},
\end{equation}
Upon substitution of this expression into Eq.~(\ref{eq:zeroth_eta}) we
discover that $\eta_0'$ is an integrating factor.  Using the second
condition in Eq.~(\ref{eq:zeroth_eta_bc}) we obtain
\begin{equation}
  \label{eq:eta_0'}
  \eta_0' = \frac{1}{\tilde\alpha}    
  \left[
    2f - \tilde s^2\frac{(g(\eta_\mathrm{max}) - g)^2}{h}
  \right]^{1/2}.
\end{equation}
The first condition in Eq.~(\ref{eq:zeroth_eta_bc}) furnishes the relation
between $\eta_\mathrm{min}$ and $\eta_\mathrm{max}$ 
\begin{equation}
  \label{eq:minmax}
  g(\eta_\mathrm{max}) = g(\eta_\mathrm{min}) +
  \frac{\sqrt{2f(\eta_\mathrm{min})h(\eta_\mathrm{min})}}{\tilde s}.
\end{equation}
Armed with this condition we can obtain an equation for
$\eta_\mathrm{min}$ via
\begin{equation}
  \label{eq:etamin}
  \frac{\Delta\theta}{2} = \int_0^{\zeta_0} d\zeta \, \theta_0' =
  \tilde \alpha \tilde s \int_{\eta_\mathrm{min}}^{\eta_\mathrm{max}}
  d\eta \, \frac{g(\eta_\mathrm{max}) - g}{h\sqrt{2f  -  \tilde
      s^2 \,(g(\eta_\mathrm{max}) - g)^2/h}},
\end{equation}
where we took advantage of the monotonicity of $\eta_0$.  
Once $\eta_\mathrm{min}$ is found we can calculate the width of the
boundary 
\begin{equation}
  \label{eq:ell_0}
  \zeta_0 = \tilde \alpha \int_{\eta_\mathrm{min}}^{\eta_\mathrm{max}}
  \frac{d\eta}{\sqrt{2f  -  \tilde s^2 \,(g(\eta_\mathrm{max}) -
      g)^2/h}},
\end{equation}
and the interfacial energy
\begin{multline}
  \label{eq:gamma}
  \gamma = \tilde s \Delta\theta \, g(\eta_\mathrm{max})
  +   2\tilde\alpha \int_{\eta_\mathrm{max}}^1 d\eta \, \sqrt{2f} \\ +
  2\tilde\alpha \int_{\eta_\mathrm{min}}^{\eta_\mathrm{max}} d\eta
  \,\sqrt{2f - \tilde s^2(g(\eta_\mathrm{max}) - g)^2/h}.
\end{multline}
Remarkably, these formulas reduce the flat boundary problem to
quadratures. We are able to compute $\eta_\mathrm{min}$ and
$\eta_\mathrm{max}$ from Eqns.~(\ref{eq:minmax}) and
(\ref{eq:etamin}), the interface width from Eqn.~(\ref{eq:ell_0}), and
the surface energy from Eqn.~(\ref{eq:gamma}).

As useful as these expressions are, insight can be better obtained
from analytic expressions, since the above four equations must
typically be solved numerically. Thus, let us examine the behavior of
the zeroth order solution in certain limits in which an approximate
solution can be found.  The integral on the right hand side of
Eq.~(\ref{eq:etamin}) can be calculated approximately when it is
small.  We identify two such situations.

\subsection{Small $h$ approximation to the zeroth order solution}
\label{sec:zeroth_small_h}

To make contact with Ref.~\cite{kwc3} in which the $|\nabla\theta|^2$
term in the free energy is set to zero, we consider a limit in which
its coefficient vanishes $h \rightarrow 0$.  Expanding
Eq.~(\ref{eq:minmax}) in powers of $h$ we obtain to lowest order
\begin{equation}
  \label{eq:etamax_small_h}
  \eta_\mathrm{max} - \eta_\mathrm{min} \approx
  \frac{\sqrt{2f^\mathrm{min}h^\mathrm{min}}}{\tilde s \,
    g_\eta^\mathrm{min}} \equiv G \ll 1,
\end{equation}
where $f^\mathrm{min} \equiv f(\eta_\mathrm{min})$, etc.  Thus, as $h
\rightarrow 0$, the difference between $\eta_\mathrm{min}$ and
$\eta_\mathrm{max}$ vanishes, while they remain well separated from
$1$.  Therefore, $f$, $g$ and $h$ and their derivatives are regular at
the undistinguished point $\eta_\mathrm{min} < 1$.

Let us define an auxiliary parameter $y$ via $\eta = \eta_\mathrm{min}
+ Gy$ and expand Eq.~(\ref{eq:etamin}) in powers of $G$.  We obtain
\begin{equation}
  \label{eq:etamin_small_h}
  \frac{\Delta\theta}{2} \approx \frac{\tilde \alpha}{\tilde s}
  \frac{\sqrt{2f^\mathrm{min}}}{g_\eta^\mathrm{min}} \int_0^1 dy
  \frac{1 - y}{\sqrt{y(2 - y)}} = \frac{\tilde \alpha}{\tilde s}
  \frac{\sqrt{2f^\mathrm{min}}}{g_\eta^\mathrm{min}}.
\end{equation}
One could in principle now invert (\ref{eq:etamin_small_h}) to solve
for $\eta_\mathrm{min}$.  The width of the boundary $\zeta_0$ can be
then found with the same accuracy
\begin{equation}
  \label{eq:ell_0_small_h}
  \zeta_0 \approx \frac{\pi}{2}\frac{\tilde \alpha}{\tilde s}
  \frac{\sqrt{h^\mathrm{min}}}{g_\eta^\mathrm{min}} \rightarrow 0.
\end{equation}
The interfacial energy is, in this same approximation,
\begin{equation}
  \label{eq:gamma_small_h}
  \gamma \approx \tilde s\, g^\mathrm{min} \Delta\theta + 2\tilde \alpha
  \int^1_{\eta_\mathrm{max}} d\eta\sqrt{2f}.
\end{equation}

For a special case examined in Ref.~\cite{kwc3}, $f = \frac{1}{2}(1 -
\eta)^2$, $g = \eta^2$, and $h = 0$, Eq.~(\ref{eq:etamin_small_h}) is
easily invertible.  Indeed, the expressions for $\eta_\mathrm{min}$
and $\gamma$ coincide with those of Ref.~\cite{kwc3}.

We finally remark that within this approximation, the behavior of
$\gamma$ and $\zeta_0$ in the limit of small misorientation
$\Delta\theta$ depends on the properties of $g$ and $h$ near $\eta =
1$.  To see why this is true, let us look at
Eq.~(\ref{eq:etamin_small_h}) in the limit of small $\Delta\theta$.
Unless $g_\eta$ is singular at some value of $\eta$ other than $1$
(unphysical), small $\Delta\theta$ implies small $1 -
\eta_\mathrm{min}$ since $f(1) = 0$.  We shall see in the next
subsection that it is true in general.  This fact is not surprising
since small angle boundaries can be thought of as arrays of distant
dislocations.

\subsection{Small $\Delta\theta$ approximation to the zeroth order solution}
\label{sec:zeroth_small_dtheta}

All parameters being fixed, the right hand side of
Eq.~(\ref{eq:etamin}) can be small if and only if 
\begin{equation}
  \label{eq:rho}
  \rho \equiv \eta_\mathrm{max} - \eta_\mathrm{min} \ll 1.  
\end{equation}
Since $f(1) = 0$, it is clear from Eq.~(\ref{eq:minmax}) that 
\begin{equation}
  \label{eq:lambda}
  \lambda \equiv 1 - \eta_\mathrm{min} \ll 1.
\end{equation}
Also, since we chose $f(1) = 0$ we can approximate it by a parabola
$f(\eta) \approx \frac{1}{2} f_{\eta\eta}(1)(1 - \eta)^2$ near $\eta =
1$.  Then
\begin{equation}
  \label{eq:minmax_small_dtheta}
  \rho \, g_\eta^\mathrm{min} \approx \lambda \,
  \frac{\sqrt{f_{\eta\eta}(1) \, h^\mathrm{min}}}{\tilde s}.
\end{equation}
If the behavior of $g$ and $h$ near $\eta = 1$ is known, we can obtain
a complete solution to the zeroth order problem in this limit.
Instead let us focus on the scaling of the interface width $\zeta_0$
and the interfacial energy $\gamma$ in the $\Delta\theta \rightarrow
0$ limit.  This scaling can be deduced without a complete solution of
the zeroth order problem.

Suppose that near $\eta = 1$ ($0 < \lambda \ll 1$)
\begin{equation}
  \label{eq:near_eta=1}
  g_\eta^\mathrm{min} \sim \lambda^\beta, \quad h^\mathrm{min} \sim
  \lambda^{2\omega}.
\end{equation}
Then from (\ref{eq:minmax_small_dtheta}) we obtain
\begin{equation}
  \label{eq:rho2}
  \rho \sim \lambda^{1 + \omega - \beta}.
\end{equation}
Note that since $\rho < \lambda$ by definition, the scaling exponent
in Eq.~(\ref{eq:rho2}) must be greater than or equal to $1$.  This
means that in this limit, the zeroth order solution exists only when
$\omega \ge \beta$.

To determine the scaling of the right hand side of
Eq.~(\ref{eq:etamin}) we define a finite integration variable $y \in
[0,1]$ via $\eta - \eta_\mathrm{min} = \rho y$.  Then
\begin{subequations}
  \label{eq:etamin_small_dtheta}
  \begin{eqnarray}
    \label{eq:deta}
    d\eta & \sim & \rho \sim \lambda^{1 + \omega - \beta}, \\
    \label{eq:numerator}
    \frac{g(\eta_\mathrm{max}) - g}{h} & \sim & \rho \,
    \frac{g_\eta^\mathrm{min}}{h^\mathrm{min}} \sim \lambda^{1 - \omega}, \\
    \label{eq:denominator}
    2f  -  \tilde s^2 \,\frac{(g(\eta_\mathrm{max}) -  g)^2}{h} & = &
    \\ \nonumber
    \frac{1}{h}
    \left[
      (2fh - 2f^\mathrm{min} h^\mathrm{min}) \right.& + &
      2\tilde s \, (g - g^\mathrm{min})
      \sqrt{2f^\mathrm{min}h^\mathrm{min}} \\ \nonumber  \left. - \tilde s^2 
      (g - g^\mathrm{min})^2 
    \right] & \sim & \lambda^{2 + \beta - \omega}.
  \end{eqnarray}
\end{subequations}
We used the fact that $\omega \geq \beta$ to establish that the second
term in the square brackets of (\ref{eq:denominator}) dominates (when
$\omega = \beta$ all three terms in these square brackets are equally
important).  Substituting scaling relations
(\ref{eq:etamin_small_dtheta}) into Eq.~(\ref{eq:etamin}) we obtain
\begin{equation}
  \label{eq:lambda2}
  \Delta\theta \sim \lambda^{1 + \frac{\omega - \beta}{2}}.
\end{equation}
We are now ready to calculate the behavior of $\zeta_0$ and $\gamma$ in
the $\Delta\theta \rightarrow 0$ limit.  Substituting the scaling
relations (\ref{eq:etamin_small_dtheta}) into (\ref{eq:ell_0}) we
obtain
\begin{equation}
  \label{eq:ell_0_small_dtheta}
  \zeta_0 \sim \lambda^{\frac{3}{2}(\omega - \beta)} \sim
  (\Delta\theta)^\frac{3(\omega - \beta)}{2 + \omega - \beta}.
\end{equation}
The interfacial energy consists of three pieces $\gamma = \gamma_1 +
\gamma_2 + \gamma_3$ which scale differently with $\Delta\theta$.  We
list them separately
\begin{subequations}
  \label{eq:gamma_small_dtheta}
  \begin{eqnarray}
    \label{eq:gamma_f_bdry}
    \gamma_1 & = & 2\tilde \alpha
    \int_{\eta_\mathrm{min}}^{\eta_\mathrm{max}} d\eta 
    \, \sqrt{2f - \tilde s^2(g(\eta_\mathrm{max}) - g)^2/h} \\ \nonumber &
    \sim & \lambda^{2 + \frac{3}{2}(\omega - \beta)}, \\ 
    \label{eq:gamma_f_grain}
    \gamma_2 & = & 2\tilde \alpha \int_{\eta_\mathrm{max}}^1 d\eta
    \sqrt{2f} \sim \lambda^2 \sim 
    (\Delta\theta)^\frac{4}{2 + \omega - \beta}, \\
    \label{eq:gamma_g_bdry}
    \gamma_3 & = & \tilde s \Delta\theta \, g(\eta_\mathrm{max}) \sim
    g^\mathrm{min} \Delta\theta. 
  \end{eqnarray}
\end{subequations}
We can make several observations.  First, since $\omega \geq \beta$,
$\gamma_2$ always dominates $\gamma_1$ in the $\Delta\theta
\rightarrow 0$ limit.  Second, when $\beta < -1$ we can integrate
(\ref{eq:near_eta=1}) to obtain $g^\mathrm{min} \sim \lambda^{1 +
  \beta}$.  When $\beta = -1$, we obtain in a similar fashion
$g^\mathrm{min} \sim \ln\lambda$.  For $\beta > -1$ the behavior of $g$
near $\eta = 1$ is arbitrary.  Thus, the scaling of the interfacial
energy (whether it is dominated by $\gamma_2$ or $\gamma_3$) in the
limit of the small misorientation can be freely controlled by
adjusting the behavior of $g$ and $h$ near $\eta = 1$.

\section{First order solutions: interface mobility}
\label{sec:first}

Having completed our analysis of the zeroth order in the asymptotic
expansion, we now continue on to first order.  This order in the
expansion will yield the velocity of the interface as a function of
geometry, mobilities, and surface energy.  From classical as well as
order parameter models, we expect this motion to be by curvature
\cite{AC}, and, as we show below, this is indeed the case.

To begin, in order for us to establish the matching conditions between
the two pieces of the first order solution, let us look at the
equation for $\eta_1^{(i)}$ valid in $\zeta > \zeta_+$
\begin{equation}
  \label{eq:first^m}
  -[v\tilde\tau_\eta Q + \tilde \alpha^2 \kappa] (\eta_0^{(i)})' =
  \tilde\alpha^2 (\eta_1^{(i)})'' -
  \eta_1^{(i)}f_{\eta\eta}(\eta_0^{(i)}).
\end{equation}
Progress can be made by noticing that $\eta_1^{(i)} = (\eta_0^{(i)})'$
is a solution of (\ref{eq:first^m}) with the left hand side set to
$0$.  We can take advantage of this fact by multiplying
Eq.~(\ref{eq:first^m}) by $\eta_0^{(i)}$ and integrating over the
grain interior.  Using the matching condition with the outer solution
which states that all derivatives vanish as $\zeta \rightarrow
\infty$, we obtain
\begin{multline}
  \label{eq:first_eta_bc}
  \tilde \alpha^2
  \left[
    (\eta_1^{(i)})' (\eta_0^{(i)})' - \eta_1^{(i)} (\eta_0^{(i)})''
  \right]_{\zeta = \zeta_+} = \\ \int_{\zeta_+}^\infty d\zeta \, [v
  \tilde \tau_\eta Q + \tilde\alpha^2 \kappa][(\eta_0^{(i)})']^2.
\end{multline}
As we mentioned before, all orders of the $\epsilon$-expansion and
their derivatives must be continuous at $\zeta_+$ or, equivalently, at
$\zeta_0$.  Consequently, we may drop the $(i)$ superscript from the
expression on the left hand side of Eq.~(\ref{eq:first_eta_bc}).  This
condition will suffice to set $\eta_1$ at $\zeta = \zeta_0$.  We will
not need the boundary condition for $\theta_1$ at $\zeta = \zeta_0$.

Turning our attention to the boundary region $\zeta \in [0, \zeta_0]$
we write down the first order equations
\begin{widetext}
  \begin{subequations}
    \label{eq:first}
    \begin{eqnarray}
      \label{eq:first_eta}
      -v\tilde\tau_\eta Q \, \eta_0' & = & \tilde\alpha^2(\eta_1'' +
      \kappa \eta_0') - f_{\eta\eta} \, \eta_1 -
      \frac{h_{\eta\eta}}{2}
      \eta_1 (\theta_0')^2 - h_\eta \, \theta_0' \theta_1' - \tilde s
      g_{\eta\eta} \, \eta_1 \theta_0' - \tilde s g_\eta \, \theta_1',
      \\ \label{eq:first_theta}
      -v\tilde\tau_\theta \eta_0^2 P \, \theta_0' & = & \tilde s
      \kappa g + \kappa h\, \theta_0' + [h \, \theta_1' + h_\eta \,
      \eta_1 \theta_0' + \tilde s g_\eta \, \eta_1]',
    \end{eqnarray}
  \end{subequations}
\end{widetext}
where we used the fact that
$\nabla\cdot[\frac{\nabla\theta}{|\nabla\theta|}] = \kappa$.  The
couplings $f$, $g$, $h$, and their derivatives, and $Q$,
$P$ in Eqs.~(\ref{eq:first}) are evaluated at the zeroth order
solution.

We can fix the boundary conditions at $\zeta = 0$ for the first order
functions by noticing that $\eta_1$ is odd while $\theta_1$ is even in
$\zeta$.  Therefore
\begin{equation}
  \label{eq:first_0_bc}
  \eta_1(0) = 0, \quad \theta_1'(0) = 0.
\end{equation}
Using the second condition we integrate (\ref{eq:first_theta}) to
obtain
\begin{multline}
  \label{eq:theta_1'}
  h(\eta_0) \, \theta_1' = -[h_\eta(\eta_0) \, \theta_0 + \tilde s \,
  g_\eta(\eta_0)]\, \eta_1 \\ - \int_0^\zeta d\zeta' \,
  (v\tilde\tau_\theta \, P(\eta_0, \hat n \theta_0') \,
  \eta_0^2 \theta_0' - \tilde s \kappa \, g(\eta_0) - \kappa \, h(\eta_0) \,
  \theta_0').
\end{multline}
Upon substitution of this expression into (\ref{eq:first_eta}) we
obtain an equation for $\eta_1$ of the form 
\begin{equation}
  \label{eq:eta_1''}
  \mathcal{L}[\eta_1] \equiv \tilde\alpha^2 \eta_1'' + C(\eta_0,
  \theta_0) \eta_1 = D(\eta_0, \theta_0),
\end{equation}
where
\begin{multline}
  \label{eq:D}
  D(\eta_0, \theta_0) = -\eta_0'(v\tilde\tau_\eta Q + \kappa) \\ - 
  \frac{h_\eta \theta_0' + \tilde s g_\eta}{h} \int_0^\zeta d\zeta' \,
  [v\tilde\tau_\theta \, P \, \eta_0^2\theta_0' + \kappa (h
  \theta_0' + \tilde s g)].
\end{multline}
The exact form of $C$ is unimportant since we can show by direct
substitution that (as in a conventional asymptotic expansion problem)
$\mathcal{L}[\eta_0'] = 0$.  This fact can be utilized to obtain a
solvability condition which determines the velocity of the interface
$v$.  Multiplying Eq.~(\ref{eq:eta_1''}) by $\eta_0'$ and integrating
over $[0, \zeta_0]$ we obtain
\begin{multline}
  \label{eq:solvability}
  \tilde\alpha^2
  \left[
    \eta_0' \eta_1' - \eta_0'' \eta_1
  \right]_0^{\zeta_0} = \tilde \alpha^2
  \left[
    (\eta_1^{(i)})' (\eta_0^{(i)})' - \eta_1^{(i)} (\eta_0^{(i)})''
  \right]_{\zeta = \zeta_0}  \\
  = \int_{\zeta_0}^\infty d\zeta \, [v \tilde \tau_\eta \, Q +
  \tilde\alpha^2 \kappa][(\eta_0^{(i)})']^2 \\
  = - \int_0^{\zeta_0} d\zeta \, [v \tilde \tau_\eta \, Q +
  \tilde\alpha^2 \kappa](\eta_0')^2 \\ 
   -\int_0^{\zeta_0} d\zeta\, \theta_0'[v\tilde\tau_\theta \, P
  \, \eta_0^2 \theta_0' + \kappa(h \theta_0' + \tilde s g)].
\end{multline}
Solving for the interface velocity $v$ and using
Eqs.~(\ref{eq:theta_0'}), (\ref{eq:eta_0'}) and (\ref{eq:gamma})
yields
\begin{equation}
  \label{eq:v}
  v = - \kappa\gamma\mathcal{M}
\end{equation}
As expected, $v$ is proportional to both the curvature and the energy
of the interface.  The mobility is given by
\begin{equation}
  \label{eq:M}
  \frac{1}{\mathcal{M}} = \tilde\tau_\eta \int_0^\infty d\zeta \,
  Q(\eta_0, \hat n \theta_0') (\eta_0')^2 + \tilde\tau_\theta
  \int_0^\infty d\zeta \, P(\eta_0, \hat n \theta_0') \eta_0^2
  (\theta_0')^2. 
\end{equation}
We dropped the superscript $(i)$ since it is clear that solutions
which are valid in the grain interior should be used in (\ref{eq:M})
for $\zeta > \zeta_0$.

Just as in a conventional formal asymptotic analysis, we obtained the
normal velocity of the interface without having to solve the first
order equations.  This is a general feature of analyses of this kind
and allows one to express the mobility of the interface in terms of
the properties of a stationary interface.

To better understand the behavior of the interface mobility
$\mathcal{M}$, let us apply the approximations of sections
\ref{sec:zeroth_small_h} and \ref{sec:zeroth_small_dtheta} to
Eq.~(\ref{eq:M}).

\subsection{Mobility in the small $h$ limit}
\label{sec:first_small_h}

To illustrate the importance of the $|\nabla\theta|^2$ term in the
free energy for the motion of boundaries, let us again consider the
limit in which its coefficient $h$ vanishes.  The second term in
Eq.~(\ref{eq:M}) dominates in this limit.  Assuming for the sake of
the argument that $P$ does not depend on $\nabla\theta$, we obtain
\begin{equation}
  \label{eq:M_small_h}
  \frac{1}{\mathcal{M}} \approx \frac{\pi \, \tilde\tau_\theta \,
  \tilde\alpha}{2\tilde s} \, \frac{P^\mathrm{min} \eta_\mathrm{min}^2
  f^\mathrm{min}}{g_\eta^\mathrm{min} \sqrt{h^\mathrm{min}}}
  \rightarrow \infty.
\end{equation}
The mobility of the grain boundary therefore vanishes as $h^{1/2}$ in
this limit in support of our claim that the $|\nabla\theta|^2$ term is
required for migration of boundaries.

\subsection{Mobility in the small $\Delta\theta$ limit}
\label{sec:first_small_dtheta}

It is instructive to trace the behavior of the interface mobility in
the limit of the vanishing misorientation.  For simplicity we assume
here that the mobility functions $P$ and $Q$ are regular and assume a
non-zero value at $\eta = 1$ and $\nabla \theta = 0$.  Using
(\ref{eq:etamin_small_dtheta}) we obtain the scaling of the three
pieces which make up the right hand side of (\ref{eq:M})
\begin{subequations}
  \begin{eqnarray}
    \int_0^{\zeta_0} d\zeta \, Q (\eta_0')^2 & \sim & \lambda^{2 +
    \omega - \beta}, \\
    \int_{\zeta_0}^\infty d\zeta \, Q (\eta_0')^2 & \sim & \lambda^{2
      + \frac{\beta - \omega}{2}}, \\
    \int_0^{\zeta_0} d\zeta\, P \, \eta_0^2 (\theta_0')^2 & \sim &
      \lambda^{2 - \frac{\omega + 3\beta}{2}}.
  \end{eqnarray}
\end{subequations}
Eq.~(\ref{eq:lambda2}) can now be used to determine the behavior of
$\mathcal{M}$ in the $\Delta\theta \rightarrow 0$ limit.  We obtain
\begin{equation}
  \label{eq:M_small_dtheta}
  \frac{1}{\mathcal{M}} \sim 
  \begin{cases}
    (\Delta\theta)^{\frac{4 - \omega - 3\beta}{2 + \omega - \beta}},
    & \beta > 0, \cr
    (\Delta\theta)^{\frac{4 - \omega + \beta}{2 + \omega - \beta}}, &
    \beta \le 0.
  \end{cases}
\end{equation}
To illustrate, if $g$ and $h$ are regular at $\eta = 1$ so that
$\omega = \beta = 0$ our analysis yields $\gamma \sim \Delta\theta$
while $\mathcal{M}^{-1} \sim (\Delta\theta)^2$ so that the interface
velocity diverges in the limit of vanishing misorientation $v \sim
(\Delta\theta)^{-1}$.

\section{Grain rotation}
\label{sec:rotation}

As we discussed above, this model also allows for the grains to
rotate. Let us consider a fully developed grain structure.  In the sharp
interface limit, the solution consists of a set of regions $G_i$ of
spatially uniform $\theta$ (grains) separated by narrow (or order
$\epsilon$) grain boundaries.  To calculate the rotation rate
$\Omega_i$ of grain $i$, we integrate Eq.~(\ref{eq:theta_eq}) over the
grain interior $G_i$.  We obtain
\begin{multline}
  \label{eq:Omega_i}
  \tau_\theta \Omega_i \int_{G_i} dA \ \eta^2 P(\eta, 0)
  = s \oint_{\partial G_i} d\sigma \, g(\eta)
  \ \hat n \cdot \left[\frac{\nabla \theta}{|\nabla\theta|}\right]
  \\ \approx s \sum_j \zeta_{ij} \, g(\eta_{\mathrm{max}}^{ij}) \,
  \mathrm{sign}(\Delta\theta_{ij}).
\end{multline}
Here $\zeta_{ij}$ is the length of the common boundary between grains
$i$ and $j$.  The summation is over the neighboring grains $j$ of
orientation $\theta_j$.  In integrating by parts we used the
continuity of $\mathbf{b} \equiv \frac{\nabla\theta} {|\nabla\theta|}$
at the edge of the grain \cite{kg99}.  In the grain boundary $S$,
$\mathbf{b}$ is a unit vector in the direction of increasing $\theta$.
Thus, away from a triple junction, $\mathbf{b} = \hat n_{ij} \,
\mathrm{sign} (\Delta\theta_{ij})$ on both edges
$\Gamma_{\pm}^\epsilon$ of the boundary between grains $i$ and $j$.
The unit normal $\hat n_{ij}$ points from grain $i$ into grain $j$.
The periodicity of $\theta$ must be taken into account to calculate
$\mathrm{sign}(\Delta\theta_{ij})$.

Recall that $\eta$ is exponentially close to $1$ in all of $G_i$
except a narrow (of order $\epsilon$) strip near the boundary.
Therefore, if $\eta = 1$ is neither a zero nor a singularity of
$P$, the integral on the left hand side of (\ref{eq:Omega_i})
is approximately equal to $A_i \, P(1,0)$, where $A_i$ is the
area of the $i$-th grain.  This choice of $P$ is problematic,
however, since in the sharp interface limit $\Omega_i \sim
s/\tau_\theta \sim 1/\epsilon$.  We therefore reach a remarkable
conclusion that when $P$ is regular at $\eta = 1$, grain
rotation dominates grain boundary migration in the sharp interface
limit.  The reason for this unexpected result is that
$\partial\theta/\partial t$ is continuous across the edge of the grain
boundary.  Inside the grain boundary, this time derivative must scale
with the inverse of the grain boundary width $1/\epsilon$.  The time
rate of change of the orientation $\theta$ in the grain interior must
therefore have the same scaling.

Another way of interpreting the divergence of the rotation rate in the
sharp interface limit is to consider what happens in the interior of
the grain.  Since the orientation order parameter is constrained to
remain uniform in space, it may be thought of as obeying a diffusion
equation with infinite diffusivity.  It is therefore not surprising
that for a generic choice of the mobility function $P$, the rotation
rate is of the same order as the rate of change of $\theta$ in the
grain boundaries (fast).  The resolution of this paradox is in fact
satisfyingly physical.  When material is nearly a perfect crystal
($\eta$ close to 1), i.e.\ there are few defects, one should expect
the rate at which the order parameters change to vanish.  This can be
accomplished by letting the mobility functions $P$ and $Q$ diverge at
$\eta = 1$.  

In fact, that for a particular choice of weakly singular $P(\eta, 0)
\sim -\ln(1 - \eta)$ the grain rotation rate no longer diverges in the
sharp interface limit.  The integral on the left hand side of
(\ref{eq:Omega_i}) may be calculated approximately in the sharp
interface limit.  We will need an expression for $\eta_0^{(i)}$ far
away from the boundary where it is close to $1$.  Using the
approximation for $f \approx \frac{1}{2}f_{\eta\eta}(1)(1 - \eta)^2$,
we obtain
\begin{equation}
  \label{eq:eta_inside}
  1 - \eta_0^{(i)} \sim \exp
  \left(
    -\frac{\sqrt{f_{\eta\eta}(1)}}{\tilde\alpha}\zeta
  \right).
\end{equation}
Therefore, focusing on the $\epsilon$-scaling in the sharp interface
limit, we deduce that
\begin{equation}
  \label{eq:singular_Q2}
  \int_{G_i} dA \ \eta^2 \ln(1 - \eta) \sim L\epsilon \int^{L/\epsilon}
  \zeta \, d\zeta \sim \frac{L^3}{\epsilon},
\end{equation}
where $L$ is some macroscopic length of the order of the grain size.
Therefore $\Omega_i \sim s\epsilon/\tau_\theta \sim 1$ no longer
diverges in the sharp interface limit.  Another important consequence
of this argument is that since the perimeter of a grain is
proportional to $L$, the rotation rate $\Omega$ is inversely
proportional to $L^2$
\begin{equation}
  \label{eq:Omega}
  \Omega \sim \frac{1}{L^2}.
\end{equation}
This prediction is independent of the choice of other the model
functions.  It is inconsistent with the heuristic derivations
\cite{gessinger,chen_balluffi} of the rotation rate due to the
diffusion of atoms along the grain boundary.  These studies obtain a
$1/L^3$ or $1/L^4$ scaling of $\Omega$ depending on the mechanism.  In
a separate study, Martin \cite{Martin} assumed that rotation is caused
by viscous motion of dislocations and obtained a rotation rate which
was independent of $L$.

\section{Approximate model of a single circular grain}
\label{sec:toy}

To illustrate the predictions of the sharp interface limit calculation
of the preceding sections, we consider a circular grain of radius $R$
and orientation $\theta$ embedded in an immovable matrix of
orientation $0$.  To make analytical progress we choose $Q = 1$, $P =
-\ln(1 - \eta)$, $f = \frac{1}{2}(1 - \eta)^2$, $g = -2\eta - 2\ln(1 -
\eta)$, $h = 1$ to ensure finite rotation rate in the sharp interface
limit and Read-Shockley \cite{readshockley} behavior low angle
boundaries.  We also restrict ourselves to the small $\theta$
approximation.
\begin{figure}[htbp]
  \begin{center}
    \includegraphics[width=2.5in]{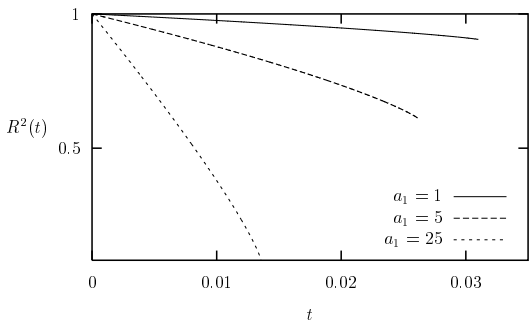}
    \caption{Evolution of the squared radius $R^2(t)$ of the circular
      grain for $a_2 = 1$ and three values of $a_1$.}
    \label{fig:R}
  \end{center}
\end{figure}
\begin{figure}[htbp]
  \begin{center}
    \includegraphics[width=2.5in]{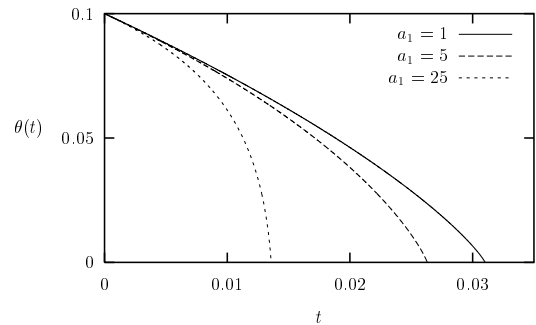}
    \caption{Evolution of the orientation $\theta(t)$ of the circular
      grain for the same three values of $a_1$.}
    \label{fig:theta}
  \end{center}
\end{figure}
In this limit we can carry out the expansion in detail to obtain
\begin{equation}
  \label{eq:gamma_M_toy}
  \gamma \approx \tilde s \, \theta
  \left(
    1 - \ln\frac{\tilde s \theta}{\tilde\alpha}
  \right), \quad \zeta_0 \approx \frac{\pi}{4} \sqrt{\frac{\tilde
  \alpha\theta}{\tilde s}}, \quad
  \frac{1}{\mathcal{M}} \approx \frac{\tilde \tau_\eta \tilde s \,
    \theta}{2\tilde\alpha^2}.
\end{equation}
Applying the motion by curvature result (\ref{eq:v}) we obtain
\begin{equation}
  \label{eq:R_dot}
  2R \dot R \approx a_1\ln\theta,
\end{equation}
where $a_1 = 4\tilde\alpha^2/\tilde\tau_\eta$.  The expression for the
rotation rate with the radius of the grain $R$ and misorientation
$\theta$ can be obtained via the arguments of this section above.  We
obtain
\begin{equation}
  \label{eq:theta_dot}
  \dot \theta \approx a_2 \frac{\ln \theta}{R^2},
\end{equation}
where $a_2 = 6\tilde s \tilde \alpha/\tilde \tau_\theta$.  Solutions
to these equations for $R(0) = 1$ and $\theta(0) = 0.1$ are given in
Figs.~\ref{fig:R} and \ref{fig:theta}.  The ratio $a_1/a_2$ controls
the behavior of the solution.  When this ratio is small, rotation
dominates the dynamics so that the radius of the grain is not
significantly reduced by the time the grain rotates into alignment
with the matrix.  On the other hand, when this ratio is large, the
evolution of the radius squared of the grain is almost linear in time
as in the case of the motion by curvature.

\section{Discussion}
\label{sec:discussion}

In this paper we analyze a modified version of the phase field model
of KWC \cite{kwc}. This model is constructed to describe rotation of
crystalline grains coupled to the motion of grain boundaries.  The
order parameter $\theta$ reflects the local crystal orientation,
whereas $\eta$ represents local crystalline order.  The
Ginzburg-Landau free energy depends only on $\nabla\theta$ and is
therefore invariant under rotations.  Inclusion of the non-analytic
$|\nabla\theta|$ term into the free energy results in singular
gradient flow equations.  However, this singularity can be dealt with
in a systematic way.

Quite generally, solutions to the model represent a collection of
regions of uniform $\theta$---grains---connected by narrow (of order
$\epsilon$) internal layers---grain boundaries.  We are able to
calculate the velocity of the boundaries in the limit of vanishing
interface thickness, and find that it is proportional to the product
of surface energy, curvature of the interface, and a mobility which
depends on model parameters.  The behavior of the interfacial width,
energy and mobility in the limit of the small misorientation is
controlled by the behavior of the model couplings near $\eta = 1$.

We calculate the rate of grain rotation in the sharp interface limit
and find that it diverges unless the mobility function $P$ is singular
at $\eta = 1$.  For a logarithmic choice of this singularity, the rate
of the grain rotation is finite and non-zero in the sharp interface
limit.  We explain this mathematical requirement by noting that the
singular term in the free energy results in infinitely fast diffusion
in the interior of a grain.  Therefore, $\eta$-dependent mobility $P$
must compensate for that fact.  This logarithmically singular $P$
leads to the conclusion that the rotation rate of a grain scales with
the inverse of its area. This is a robust prediction of our model
independent of the choice of all other couplings.

For a plausible choice of model functions, motivated by the physics of
low angle grain boundaries, we derive and solve equations describing a
circular grain embedded in a matrix.  We find that, as expected, when
the scaled coefficient $\tilde s$ of the $|\nabla\theta|$ term in the
free energy is large, rotation is fast so that the radius of the grain
does not change much by the time the grain rotates into coincidence
with the matrix.  While when $\tilde s$ is small, rotation becomes
important only when the radius is significantly reduced.

We conclude by remarking that the model may be readily generalized in
a variety of ways.  We may extend the model to three dimensions, by
constructing an appropriate tensor order parameter which reflects the
symmetries of the lattice.  Anisotropy may be included by allowing the
coefficient of the $|\nabla\eta|^2$ term to depend on $\nabla\theta$.
Alternatively, the mobility functions may be made anisotropic to yield
kinetics which depend on the orientation of the boundary region.
Overall, this model will provide a foundation for a physical, yet
still relatively mathematically simple, model of grain boundary
evolution and grain rotation.

\bibliography{../curvature}

\end{document}